\documentclass[noshowpacs,aps,twocolumn]{revtex4}

\usepackage{graphicx}
\usepackage{amsmath}

\begin{document}

\author{Bogdan-Ioan Popa}
\email{bap7@ee.duke.edu}
\author{Steven A. Cummer}
\affiliation{Department of Electrical and Computer Engineering\\
Duke University, Durham, North Carolina 27708}

\title{Derivation of effective parameters of magnetic
metamaterials composed of passive resonant LC inclusions}

\date{\today}

\begin{abstract}
We determine rigorously the effective permeability of magnetic
metamaterials composed of passive particles exhibiting magnetic
resonances. The effective permeability is expressed in terms of
circuit parameters characteristic to these particles (L, C, and R) and particle geometry. The derivation takes into account the magnetic coupling between particles. We show that a bigger concentration of
particles leads to an improvement in the metamaterial performances
(such as bandwidth or loss tangent), but this effect saturates as the particles are packed more tightly. The theory is validated through
numerical simulations of physically realizable metamaterials.
\end{abstract}

\maketitle

Since the development of the first electromagnetic metamaterials based on split-ring-resonators (SRRs) and wires, it has been acknowledged that the narrow bandwidth, high dispersion, and significant loss inside them have been the main factors that limited their practical use in commercial applications. Theoretical analysis of the individual particles that compose such metamaterials have been performed before \cite{PENDRY99,MARQUES02}, and gave valuable insight on their physics. However, a complete and reliable theoretical derivation of the effective material parameters of these metamaterials in terms of their geometry has not been developed yet.

In this paper we focus on metamaterials having negative permeability, which are commonly designed and built using periodic arrays of unit cells composed of inclusions such as SRRs \cite{SHELBY01}, omega particles \cite{RAN04}, and other types of similar resonant particles \cite{ERENTOK05}. Our purpose is to derive rigorously the effective permeability of such metamaterials in terms of the circuit parameters characteristic to the composing particles (i.e. inductance, $L$, capacitance, $C$, and resistance, $R$) and the particle geometry represented by the geometry factor $F$. We will show that larger $F$ factors lead to an increase in the performance in terms of bandwidth or loss tangent. However, we will show that $F$ cannot be made arbitrarily large and reaches an upper bound as the particles are packed closer together.  Numerical simulations will be performed to validate the theoretical results.

\begin{figure}
  \begin{center}
    \includegraphics[width=8cm]{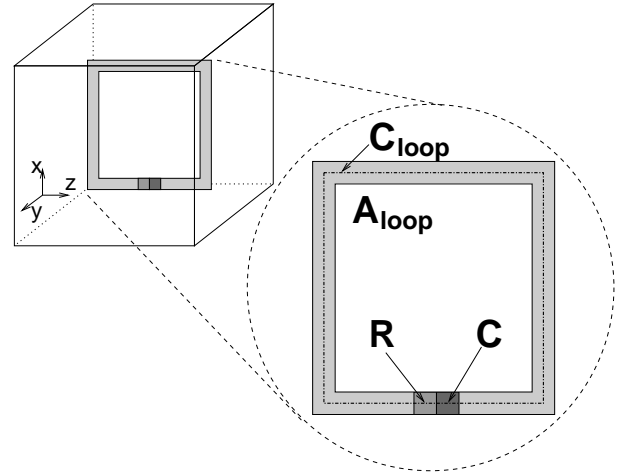}
    \caption{Unit cell containing a capacitively loaded loop. Periodic arrangements of this unit cell generate the metamaterial. Inset: The capacitively loaded loop used to generate the magnetic response of the metamaterial. The curve $C_{loop}$ follows the contour of the loop; $A_{loop}$ is the area delimited by $C_{loop}$}
    \label{fig:particle}
  \end{center}
\end{figure}

Consider the unit cell depicted in Figure \ref{fig:particle}. All the resonant particles commonly used to generate negative permeability (such as SRRs, omega particles) are essentially LC resonant circuits \cite{PENDRY99,MARQUES02} that can be represented in general by a planar, capacitively loaded loop as depicted in Figure \ref{fig:particle}.
Losses are taken into account by using a resistor, $R$, in series with the
capacitor, $C$, and the loop. For now, the loop can have an arbitrary shape.
A metamaterial is generated by using periodic arrays of this unit
cell. Consider a plane wave propagating through this metamaterial in
the $+z$ direction, such that the magnetic field,
$\mathbf{B_i}=\mathbf{\widehat{y}}B_0e^{-jk_0 z}$ ($e^{j\omega t}$
time dependence assumed), is perpendicular on the plane of the loop.
Writing Faraday's Law in integral form for an arbitrary loop inside the metamaterial we obtain:
\begin{equation} \label{eq:faraday}
\int_{C_{loop}} \mathbf{E}\cdot d\mathbf{l} = -j \omega \int_{A_{loop}}
(\mathbf{B_i}+\mathbf{B_{loop}}+\sum_n\mathbf{B_{n}})\cdot
d\mathbf{s}
\end{equation}
where $\mathbf{B_{loop}}$ is the magnetic field produced by the
current induced in the loop by the incident $\mathbf{B_i}$,
$\mathbf{B_{n}}$ is the field produced by the current induced in the
$n$-th loop. From the definition of auto inductance,
$\int_{A_{loop}}\mathbf{B_{loop}}\cdot d\mathbf{s}=L_{loop}I$, where
$L_{loop}$ is the auto-inductance of the loop, and $I$ is the current induced in the loop. Similarly, $\int_{A_{loop}}\mathbf{B_n}\cdot
d\mathbf{s}=M_n I_n$, where $M_n$ and $I_n$ are the mutual inductance between the $n$-th loop and the loop for which Eq. (\ref{eq:faraday}) is written, and, respectively, the current through the $n$-th loop. Note that $M_n$ can be positive or negative depending on the relative position between the two loops. In a bulk material, far from the
edges, because of the symmetry of the metamaterial structure, we can
consider that $I=I_n$ for every $n$. Moreover, the voltage
across the resistor in series with the capacitor is $I[R+1/(j\omega
C)]=\int_{C_{loop}} \mathbf{B_i}\cdot d\mathbf{s}$. Therefore, Eq.
(\ref{eq:faraday}) becomes:
\begin{equation} \label{eq:faraday2}
I\left(R+\dfrac{1}{j\omega C}\right) =-j\omega\left(\int_{A_{loop}}\mathbf{B_i}\cdot d\mathbf{s}+IL_{eq}\right)
\end{equation}
where $L_{eff}\equiv L+\sum_n M_n$. Letting $\omega_0\equiv
(CL_{eff})^{-1/2}$ be the resonant frequency of the loop, the above
equation gives the current induced in the loop:
\begin{equation} \label{eq:I}
I=-\dfrac{\int_{A_{loop}}\mathbf{B_i}\cdot d\mathbf{s}}{L_{eff}} \dfrac{\omega^2}{\omega^2-\omega_0^2-j\omega R/L_{eff}}
\end{equation}

In general, the magnetic flux through the loop,
$\int_{A_{loop}}\mathbf{B_i}\cdot d\mathbf{s}$, is a function of the
geometry of the loop. However, it is well known that in order for the metamaterial to be described in terms of effective material
properties, the unit cell has to be much smaller than the wavelength. Therefore, for frequencies around $\omega_0$ where the permeability of the metamaterial becomes negative, we can make the approximation
$\int_{A_{loop}}\mathbf{B_i}\cdot d\mathbf{s}\simeq B_i A_{loop}=\mu_0 H_i A_{loop}$. Note that $B_i$ and $H_i$ represent the local incident B and H fields related through $B_i=\mu_0H_i$ since the loops are placed in vacuum. For example, for square loops (i.e. a geometry widely used for building SRRs)
$\int_{A_{loop}}\mathbf{B_i}\cdot d\mathbf{s}=B_i A_{loop}
\mathrm{sinc}(k_0l_{loop}/2)$, where $l_{loop}$ is the length of the
loop. If we require $\lambda/l_{loop}=10$, as it is generally desired, then $1>\mathrm{sinc}(k_0l_{loop}/2)>0.99$, which allows us to neglect the $\mathrm{sinc}$ term. Under this approximation, it follows from (\ref{eq:I}) that the magnetic dipole moment per unit volume is given by
\begin{equation} \label{eq:M}
M=\dfrac{IA_{loop}}{V_{uc}}=H_i\dfrac{\mu_0 A_{loop}^2}{L_{eff}V_{uc}}
\dfrac{\omega^2}{\omega^2-\omega_0^2-j\omega R/L_{eff}}
\end{equation}
where $V_{uc}$ is the volume of the unit cell.

Note that $M/\omega^2$ follows a purely lorentzian model, with the quality factor of the material given by
\begin{equation} \label{eq:Q}
Q\equiv R/(L_{eff}\omega_0)=1/(\omega_0RC)
\end{equation}
For weak magnetically coupled loops, $L_{eff}\approx L$, and $Q$ is
approximately the quality factor of the individual loop. Also, note
that the term:
\begin{equation} \label{eq:F}
F\equiv\dfrac{\mu_0 A_{loop}^2}{L_{eff}V_{uc}}
\end{equation}
only depends on the geometry of the unit cell and that of the loop,
and, as recognized in Ref. \onlinecite{PENDRY99}, has an important contribution to the
performances of the metamaterial.

The relative effective permeability of the metamaterial, given by $\mu_r=1+M/H_i$, can be expressed in terms of the geometry factor, $F$, quality factor, $Q$, and resonant frequency, $\omega_0$, as:
\begin{equation} \label{eq:mur}
\mu_r=1-\dfrac{F\omega^2}{\omega^2-\omega_0^2-j\omega\omega_0/Q}
\end{equation}

\begin{figure}
  \begin{center}
    \includegraphics[width=9cm]{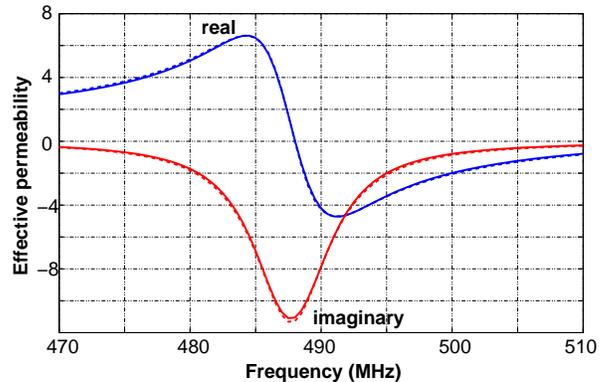}
    \caption{The effective permeability retrieved from the $S$
parameters (solid line), and the permeability computed using
(\ref{eq:mur}) (dashed line). This excellent match was obtained for
$F=0.16$ and $Q=71$.}
    \label{fig:mur}
  \end{center}
\end{figure}

To verify the accuracy of (\ref{eq:Q}) through (\ref{eq:mur}) we performed
numerical simulations in Ansoft HFSS similar to those described in Refs. \onlinecite{SMITH05} and \onlinecite{POPA05}. The metamaterial has one cubic unit cell in the direction of propagation, and has infinite extent in the transverse direction (this is simulated using perfect electric and perfect magnetic boundary conditions). The unit cell has dimensions 40 by 40 by 40 mm. The square loop has dimensions 30 by 30 mm and is loaded with a capacitor $C=1.108$ pF and a resistor $R=3.6 \Omega$. These dimensions give a resonant frequency slightly below 500 MHz, which means that the unit cell diameter is less then $\lambda/15$, and the metamaterial can be described by effective material parameters. The effective material parameters are retrieved from the $S$ parameters using a standard method widely used for these metamaterials \cite{SMITH05}, and are compared with the parameters computed theoretically
from Eq. (\ref{eq:mur}). Written in terms of capacitance, which is
assumed to be constant (the capacitive coupling between
adjacent loops can be neglected), and resonant frequency (which can be easily retrieved from the $S$ parameters), the geometry factor is (from Eq. (\ref{eq:F})) $F=\mu_0 C \omega_0^2 A_{loop}^2/V_{uc}$. We obtain $\omega_0=2\pi \cdot 487.5$ MHz, and consequently the expected $F$ and $Q$ factors are $F=0.17$, and $Q=77$. The real $Q$ and $F$ values can be retrieved by matching the permeability determined from the $S$ parameters with that given by Eq. (\ref{eq:mur}). As can be seen in Figure \ref{fig:mur} an excellent match was obtained for $F=0.16$ and $Q=71$. Note that these values are within 10\% of the expected $F$ and $Q$, which validates the theoretical relations (\ref{eq:Q}) through (\ref{eq:mur}).

Equation (\ref{eq:mur}) gives the lorentzian variation expected for
these magnetic metamaterials \cite{PENDRY99}. The relative
permeability can be controlled to some extent by controlling the
three parameters that enters (\ref{eq:mur}). Thus, $\omega_0$ is
relatively easy to control by loading the loop with lumped capacitors as suggested in Ref. \onlinecite{REYNET}. The quality factor,
$Q$, directly responsible for the losses inside the metamaterial, is harder to control since it depends on the dielectric and ohmic losses \cite{GREEGOR03}. Out of the three parameters, $F$ is the easiest to control since it depends only on the geometry of the unit cell. Moreover, a closer look to Eq. (\ref{eq:mur}) reveals that a bigger $F$ factor leads to an increase in the negative permeability bandwidth and a decrease in the magnetic loss tangent, in agreement with Ref. \onlinecite{PENDRY99}. Consequently, the next step is to determine how large $F$ can be made. Note that the parameters that enter Eq. (\ref{eq:F}) that gives $F$ are dependent on each other. The area of the loop is strongly related to its inductance, and the unit cell volume. If we reduce the unit cell volume by packing the loops closer together in an attempt to increase $F$, the effective inductance of the loops would increase as a result of an increase between the mutual inductance between the loops. This suggests that $F$ cannot be made arbitrarily large.

Since it is commonly used in literature, we will focus next on metamaterials made of cubic unit cells. For the commonly used square loop $A_{loop}^2/V_{uc}=l_{loop} (l_{loop}/l_{uc})^3$, where $l_{loop}$ is the diameter of the loop, and $l_{uc}$ is the diameter of the unit cell. Obviously, the ratio $l_{loop}/l_{uc}$ is always less than unity, and typically is less than 0.8 to reduce the coupling between adjacent loops. For example, in Ref. \onlinecite{GREEGOR03} this ratio is 0.79, while in Ref. \onlinecite{SHELBY01} it is as low as 0.52. On the other hand the inductance is proportional to $\mu_0 l_{loop}$. Since the unit cell is cubic we can neglect the inductive coupling between loops. Under this approximation, a more exact equation written for square loops made of cylindrical wires gives \cite{INAN}
\begin{equation}
L_{loop}=\dfrac{\mu_0 l_{loop}}{2\pi}\left(2.303\log_{10}\dfrac{32l_{loop}}{w}-2.853\right)
\end{equation}
where $w$ is the diameter of the wire. Typically, the term $l_{loop}/w$ is of the order of hundreds, and has a moderate influence on the inductance $L_{loop}$ because the logarithm is a slowly varying function for large arguments. Assuming $l_{loop}/w>100$ it follows that $L_{loop}>3.33\mu_0l_{loop}$, which combined to the constraint on $A_{loop}^2/V_{uc}$ gives (see Eq. (\ref{eq:F}))
\begin{equation} \label{eq:Fconstraint}
F\approx \dfrac{0.8^3\mu_0 l_{loop}}{3.33 \mu_0 l_{loop}}=0.15
\end{equation}

The above value was obtained for cubic unit cells and square loops. We expect loops of different shapes to give slightly lower values of $F$ because the ratio $A_{loop}^2/V_{uc}$ is smaller than for square loops while the inductance remains proportional to $\mu_0 \sqrt{A_{loop}}$. For example, for circular loops it can be shown in a similar manner that $F\approx 0.12$. It follows from the above analysis that for cubic unit cells and arbitrary loop shapes $F<0.2$.

This bound on $F$ is determined directly from Eq. (\ref{eq:F}), and is supported by both the parameters retrieved from the HFSS simulation (shown in Figure \ref{fig:mur}), and the results obtained by others using numerical methods \cite{SMITH02}. Note that this analytically derived upper bound on $F$ is significantly smaller than that predicted by other analytical models \cite{PENDRY99}.

\begin{figure}
  \begin{center}
    \includegraphics[width=9cm]{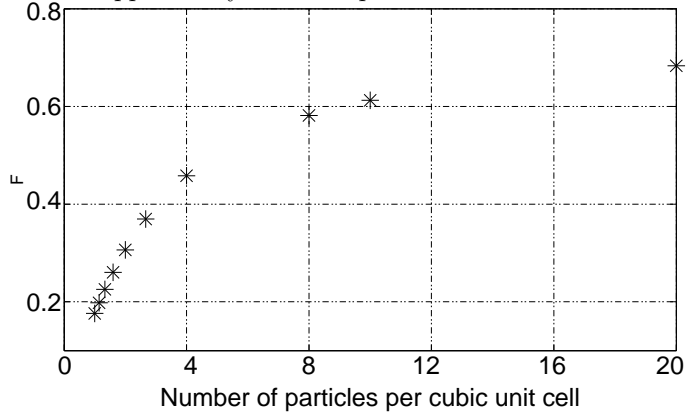}
    \caption{The geometry factor vs the concentration of particles}
    \label{fig:FvsVuc}
  \end{center}
\end{figure}

These results were obtained for cubic unit cells which are
commonly used in the design of magnetic metamaterials. To increase $F$ over the limit determined above, we can decrease the unit cell volume by reducing the cell axial dimension, or, equivalently, by increasing the concentration of loops. However, besides a decrease in $V_{uc}$, we also expect a substantial increase in the effective inductance, $L_{eff}$, due to the magnetic coupling between the loops (recall that $F\propto 1/L_{eff}$). The $F$ factor can, thus, be accurately computed either analytically, from (\ref{eq:F}) by computing the mutual inductances that give $L_{eff}$, or numerically using Ansoft HFSS and the procedure described above. For simplicity, we used the latter method to find $F$ versus the concentration of loops (expressed as the number of loops per cubic unit cell). The results are presented in Figure \ref{fig:FvsVuc}, and show the geometry factor asymptotically approaching an upper bound. Thus, even for tightly packed loops (i.e. 20 loops per cubic unit cell), $F$ remains below 0.7.

In conclusion, we derived the analytical equations that allow an accurate prediction of the effective permeability of magnetic metamaterials composed of passive resonant inclusions commonly used in literature, such as the split-ring-resonator, or the omega particle. The analysis took into consideration the magnetic coupling between these particles. We showed that, for the commonly used cubic unit cell, the geometry factor $F$ is typically 0.15. An increase in the $F$ factor can be achieved by increasing the concentration of particles. However, even for tightly packed particles $F$ reaches an upper bound significantly lower than 1. We validated these results through numerical simulations performed in Ansoft HFSS.

\end{document}